# Mapping local optical densities of states in silicon photonic structures with nanoscale electron spectroscopy

Judy J. Cha[1†], Zongfu Yu[2], Eric Smith[3], Martin Couillard[1‡], Shanhui Fan[2], David A. Muller[1]

[1]School of Applied and Engineering Physics, Cornell University, Ithaca, NY, 14853

[2]Department of Electrical Engineering, Stanford University, Stanford, CA, 94305

[3]Department of Physics, Columbia University, New York, NY, 10027

[†]Current address: Department of Materials Science and Engineering, Stanford University, Stanford, CA, 94305

[‡]Current address: Department of Materials Science and Engineering and Canadian Centre for Electron Microscopy, McMaster University, Hamilton, ON, Canada, L8S 4L7

*Relativistic electrons in a structured medium generate radiative losses such as Cherenkov and transition radiation that act as a virtual light source, coupling to the photonic densities of states. The effect is most pronounced when the imaginary part of the dielectric function is zero, a regime where in a non-retarded treatment no loss or coupling can occur. Maps of the resultant energy losses as a sub-5nm electron probe scans across finite waveguide structures reveal spatial distributions of optical modes in a spectral domain ranging from near-infrared to far ultraviolet.*

In a non-retarded, Poisson's formulation of electrodynamics, a swift electron traveling in a straight line cannot radiate. Instead, the time-varying electric field generated by the electron is evanescent, thus energy transfer to the medium can only occur in the near field, and only if the imaginary part of the medium's dielectric function, Im[$\varepsilon(\omega)$], is non-zero. In this regime, energy losses of the swift electron in the medium are used to probe bulk and surface plasmons, as well as chemically-specific core electron excitations [1-4]. However, recent theory has recognized that additional losses from relativistic electrons can provide useful information about photonic structures [5]. When the retardation is considered solving the full Maxwell's equations, the swift electron induces a time-varying polarization in the medium. When the time-dependent polarizations add coherently, radiation in the vicinity of the swift electron becomes possible – Cherenkov and Transition radiation are two well-known examples [6, 7]. In contrast to the non-retarded treatment where the loss is derived from and proportional to Im[$\varepsilon(\omega)$], the retarded loss mechanisms are most noticeable when Im[$\varepsilon(\omega)$]$\ll$1 and the non-retarded contributions are negligible. Even for mildly-relativistic electrons, these effects are large enough to provide a useful, high-spatial resolution probe of photonic nanostructures.

Spatially-resolved electron energy-loss spectroscopy (EELS) as a probe for local photonic density of states (LDOS) is attractive because of its spatial resolution of 0.1 to 5 nm. Conventional optics-based techniques are diffraction limited, resulting in spatial resolution on the order of the probing wavelength, typically a few hundred nanometers. Scanning near-field optical microscopy (SNOM) [8] overcomes the diffraction limit, but still gives a spatial resolution of tens of nanometers. While SNOM is a surface-sensitive technique, EELS probes the projected DOS of features embedded inside the photonic crystals as electrons travel through samples. Also, EELS measures the DOS principally along the direction of the electron trajectory [5] whereas SNOM measures the DOS principally on the plane of the surface over which it scans [9].

Probing photonic modes with EELS is possible because of energy losses to radiation generated by relativistic electrons in the retarded regime of electrodynamics[5], such as Cherenkov [6, 7, 10] and transition radiation [7, 11, 12]. Both types of radiation can be considered as virtual light sources due to their broad spectrum. For Si, such radiation is expected below the direct electronic transition at 3.4 eV and into the near-infrared range (a region where little energy losses are expected in a non-retarded treatment). The virtual broadband light couples to photonic DOS, in principle allowing EELS to probe photonic structures [13]. With recent improvements in monochromators and system stability, the energy resolution of EELS has improved to reach 100 meV that is comparable to ~30 nm spectral resolution in the visible frequencies. The nanometer resolution of EELS in principle allows a direct mapping of the optical DOS of radiative modes of unconventional dielectric systems that may contain defects or extend to only a few hundred nanometers. To date, the higher energy, and hence more easily resolved, non-radiative plasmonic modes have been studied extensively, such as MgO cubes [4], Si nano-particles in $SiO_2$ [10] and Ag platelets [2].

In this Letter, we map the radiative, photonic LDOS of finite optical structures using monochromated EELS at nanometer resolution. By conducting measurements in a regime where $Im[\varepsilon(\omega)] \sim 0$, the plasmonic modes are suppressed and no energy loss features, either collective or single particle, are predicted by a non-retarded treatment. This is a region where the only surviving loss mechanisms are predicted to be purely due to retardation effects, and with the slow energy dependence of $Re[\varepsilon(\omega)]$, the loss features should generally be readily recognizable as projections of the local photonic DOS [5].

(Some counterexamples can be found in ref[18], and we will discuss the conditions under which the connection between EELS and the LDOS can be made).

Two-dimensional (2D), finite photonic structures of slabs thicknesses between 1 and 1.5 μm were fabricated using a focused ion beam (30 kV voltage and 9.7 pA current). Monochromated, 200 keV electrons at 110 - 140 meV energy resolution were used to study the photonic structures. Using EEL spectroscopic images (SIs), spatial distributions of photonic modes were mapped with a 2-4 nm wide electron beam. For EELS calculations, local dielectric theories [3, 14] and the MIT photonic-bands (MPB) package [15] were used.

Thick Si slabs were first investigated to demonstrate that radiative losses could be detected by 200 keV incident electrons. Figure 1 shows experimental and calculated EELS of Si slabs; $\varepsilon(\omega)$ for Si was taken from optical data [16]. The calculated EELS in a non-retarded treatment (dotted line in Fig. 1(a)) shows no signal below 3.4 eV, the direct electronic transition of Si. However the experimental spectrum shows significant intensity below 3.4 eV that cannot be accounted for by the tails of the zero-loss beam. A calculated EELS of a 160 nm-thick Si slab using a retarded treatment (solid line in Fig. 1(a)) shows a good agreement with the experiment, confirming that the intensity below 3.4 eV is due to losses to emitted radiation. Plasmon losses of Si do not contribute to the intensity in this region, instead lying in the range of 6-20 eV, well above the energy range of interest. Fig. 1(b) shows that the emitted radiation is thickness-dependent, indicating that the radiation is guided inside the slab by the slab geometry [10].

If boundary conditions are imposed on the emitted radiation, only those modes whose frequencies satisfy the boundary conditions will be supported. These eigenmodes will appear as sharp peaks in EELS. To observe the sensitivity of EELS to these eigenmodes, 2D photonic structures were investigated. Figure 2 (a-c) show Si photonic structures with $r/a$ = 0.20, 0.27, and 0.34 where $r$ is the radius of the cylindrical holes and $a$ is the distance between two holes. For all three structures, $a$ was ~ 447 nm. We varied $r$ to observe how the photonic modes would shift in frequencies with different $r/a$ ratios.

Figure 2(d) shows simulated photonic transverse-magnetic (TM) bands of the structure shown in Fig. 2(c) [15]. Only the TM bands were calculated because EELS is more sensitive to TM modes than transverse-electric modes [10, 12, 13]. Figure 2(e) shows an EEL spectrum taken in aloof mode (beam is in vacuum near a surface) from the same structure and a simulated photonic DOS, obtained by integrating the TM bands sampled over many k-points. For simulation, the structure was assumed to be translationally invariant along the beam direction and to have infinite lateral extent, which reduced the simulation time. It should be noted that EELS shown here are momentum(k)-averaged, thus do not provide k(angle)-resolved dispersion maps of TM modes. In a scanning transmission electron microscope, angular information is limited by the convergence angle of the electron beam. Angle-resolved dispersion maps in EELS is possible [17], but the spatial resolution will be compromised.

Three distinct modes, at ~ 0.5 eV, 0.94 eV and 1.26 eV, were observed experimentally while the simulated photonic DOS show five peaks that correspond to TM bands 1 to 5, indicated by the shaded boxes that connect Fig. 2(d) and 2(e). TM bands 2 and 4, indicated by the arrows, were not resolved experimentally principally due to the lower experimental energy resolution. When the DOS was binned at the experimental energy resolution of ~ 0.1 eV, they could not be resolved either in the simulated DOS.

Some weak bumps above 1.5 eV in EELS correspond to overlapping higher energy bands. It should be noted that the simulated photonic DOS was multiplied by the evanescent decay of the time-varying electric field, $[K_0(bnE/\gamma\hbar c)]^2$ where $K_0$ is the zeroth-order modified Bessel function of the second kind, $b$ is the impact parameter, $n$ is the index of refraction of the material, E is the energy loss, $\gamma$ is the relativistic factor, $\hbar$ is the Planck's constant, and $c$ is the speed of light in vacuum [7]. For the simulation, $b$ was 70 nm and $n$ was 3.44.

The conditions under which a connection between the LDOS and EELS can be made are still being debated [5, 18]. Here, we show good agreement between the simulated DOS and the experiment, which was possible due to the translational invariance of our structures along the beam direction, z. In EELS experiments, the electron passes through the sample so the LDOS measured by EELS must always be a projection along z (i.e. integrated in some manner along z). When the z-component of the LDOS is translationally invariant along z for length scales larger than $\gamma v/\omega$ (where $\gamma$ is the relativistic factor, $v$ is the speed of the electron and $\omega$ is the mode frequencies), EELS and the z-integrated LDOS will tend to the same functional form. For the 1 – 2 eV energy range we investigate here, the thickness of the photonic structures (1 to 1.5 μm) is much larger than $\gamma v/\omega$ (~ 0.1 μm), so that we can safely compare EELS to the simulated DOS.

Figure 2(f) shows EELS acquired from the central cylinder of Si photonic structures of various *r/a* ratios in aloof mode. As expected, the characteristic mode frequencies shifted when *r/a* was varied. Two modes (TM bands 3 and 5) were examined carefully. The hollow and solid dots in Fig. 2(f) inset correspond to the energy losses of the two modes observed experimentally. The solid and dotted lines correspond to the simulated energy losses of the two modes, which agree well with the experiments.

A significant advantage of EELS for probing the photonic LDOS is the nanometer lateral spatial resolution, which allows a direct mapping of the projected spatial distributions of photonic modes in finite structures. A Si photonic structure with hollow squares (Fig. 3(a)) was fabricated to probe the spatial distribution of photonic modes. To map the spatial distribution of the modes, a 64 x 64 pixel SI was acquired from the central square. Spectra at each point of the SI contain a zero-loss peak for energy calibration, allowing tracking of the energy drifts. Three EEL spectra from the SI is shown in Fig. 3(b); the spectra were deconvolved with the zero-loss peak to increase the clarity of the modes [19].

From the photonic TM mode simulation, two peaks, at ~ 0.5 eV and ~ 0.9 eV in Fig. 3(b), are identified to be TM mode 2 and 6. The mode at ~ 0.5 eV exhibits higher intensity at the corners of the square (the top and bottom spectra), while the mode at ~ 0.9 eV exhibits higher intensity at the side of the square (spectrum in the middle). The intensity under each mode was integrated to create intensity maps (Fig. 3(c) and 3(d)). Figure 3(e) and 3(f) are simulated H-field intensity distributions of TM mode 2 and 6, averaged in the reciprocal space along the high symmetry lines ($\Gamma \rightarrow M \rightarrow R \rightarrow \Gamma$). We see a qualitative agreement in the intensity distributions of the two modes between the experiment and the simulation. Line profiles, obtained from the intensity maps, Fig. 3(c, d), are shown in Fig. 3(g) and 3(h), displaying the corner and the side modes clearly.

Sub-wavelength details of a localized defect mode or edge effects due to the finite size of photonic structures can be directly resolved with EELS at nanoscale. In photonic structures with periodicity extending over a finite region, it is interesting to observe how the photonic modes change near the edge of the structures. To observe this, we acquired 64 x 64 pixel SIs of three squares labeled square A, B, and C in Fig. 4(a). Figure 4(b) shows the spectra, summed over the pixels that make up the top side of the squares from the SIs. For square A, the modes at 0.5 eV and 0.9 eV are expected on all sides. However, for squares B and C, the modes are expected to damp out at sides where the periodicity of the photonic structure stops. As expected, the photonic modes damp out at the top side of square B (Fig. 4(b)). Fig. 4(c) and (d) show that the symmetry of the corner and the side mode is lost in square B compared to square A.

In summary, the coupling between the emitted radiation due to relativistic electrons and the photonic DOS allows for a direct electron spectroscopic investigation of optical modes with a nanoscale precision. Using monochromated EELS, multiple optical modes were detected in Si photonic structures. A key strength of EELS for probing photonic structures is its capability to probe photonic modes inside the photonic structures with the nm-scale spatial resolution. By acquiring SIs, distinct spatial distributions of photonic modes were observed. Potential applications of the technique include studying localized linear and point defect modes in photonic crystals of unconventional dielectric materials.

This work is supported by IRG 1 – Cornell Center for Materials Research (CCMR). We acknowledge the use of CCMR electron microscopy facilities supported by NSF-MRSEC (DMR No. 0520404). We thank A. Yurtsever and M. Kociak for useful discussions.

**Fig. 1.** Experimental and theoretical EELS of Si slabs. The solid and dotted lines in inset in (a) are simulated EELS of a 160 nm-thick Si slab in retarded and non-retarded regime. Plasmon losses are well above 6 eV, indicated by the bulk plasmon loss at 17 eV and the surface plasmon losses at ~7-9 eV. The features in EELS below 3.4 eV are due to radiative losses and are missing in the non-retarded treatment. (b) shows that the emitted radiation is guided as a function of the slab thickness. The thicknesses of the slab were obtained by fitting the experiment to simulation using Kröger's formalism. The spectra in (b) are normalized to the bulk plasmon and displayed with an offset for visual clarity.

**Fig. 2.** (color online). Si photonic structures of hollow cylinders. (a-c) show photonic structures with *r/a* = 0.20, 0.27, and 0.34 respectively. All scale bars are 500 nm. (d) shows the photonic TM band calculation of the structure shown in (c). The bands were traced from Γ(0,0) → M(0,0.5) → K(-1/3,1/3) → Γ(0,0) in the reciprocal space. (e) shows an EEL spectrum taken from the central cylinder of the structure with *r/a* = 0.34 in aloof mode and a simulated photonic DOS. (f) shows shifts in the photonic modes when *r/a* is varied. The solid, the short dotted, and the long dotted lines are EELS taken from *r/a* = 0.34, 0.27, and 0.2 respectively. The vertical lines (in red) are added as a visual guide to indicate the energy shifts of the modes. The inset in (f) shows the agreement between the experimentally observed shifts in the photonic modes (filled and open circles) and the simulated photonic DOS (solid and dotted lines).

**Fig. 3.** (color online). Spectroscopic image (SI) showing spatial distribution of photonic modes. (a) shows the structure where the squares are 600 nm wide and the lattice constant is 800 nm. The scale bar is 500 nm. The 64x64 pixel SI was taken from the central square. Spectra near the top side of the square are selectively shown in (b). Spectra are deconvolved with the zero-loss peak, acquired in vacuum under the same experimental conditions, and are displayed with an offset for visual clarity. The area under the peaks at 0.5 eV and 0.9 eV is integrated to create intensity maps, shown in (c) and (d). (e) and (f) show the 32x32 pixel calculated field intensity distributions of TM modes 2 and 6. The dotted boxes in (e) and (f) mark the boundary between vacuum and the Si structure. (g) and (h) show intensity line profiles averaged from the dotted boxes in (c) and (d) respectively.

**Fig. 4.** (color online). Damped photonic modes near the edge of the photonic structure. (a) shows the Si photonic structure. The scale bar is 500 nm. (b) shows EELS taken from the top side of squares labeled A, B, and C. Spectrum in solid black is from square A; spectrum in short, dotted red is from square B; and spectrum in long, dotted blue is from square C. The spectra are displayed with an offset for visual clarity and they are not deconvolved with the zero-loss peak. The intensity under the two modes at 0.5 eV and 0.9 eV (indicated by arrows in (b)) was integrated over the top side of square A and B (boxes in (a)) to create the intensity line profiles, shown in (c) and (d) respectively.

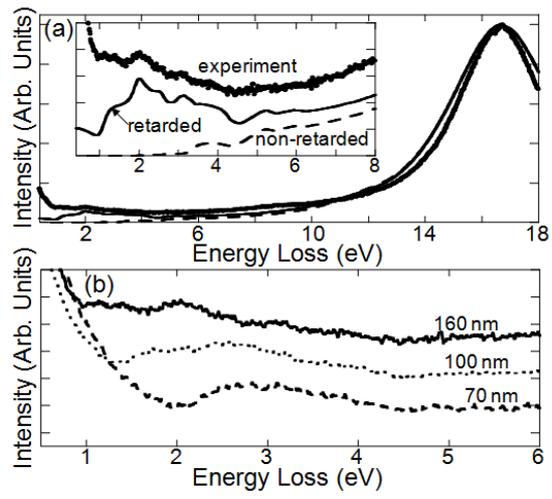

Fig. 1

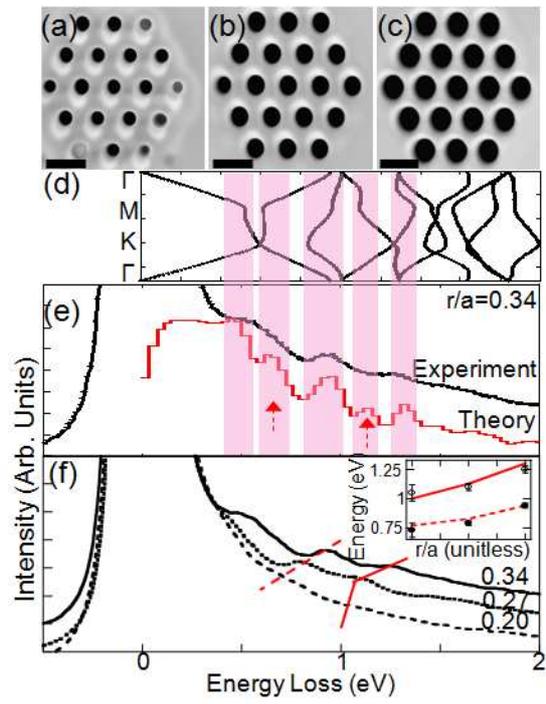

Fig. 2

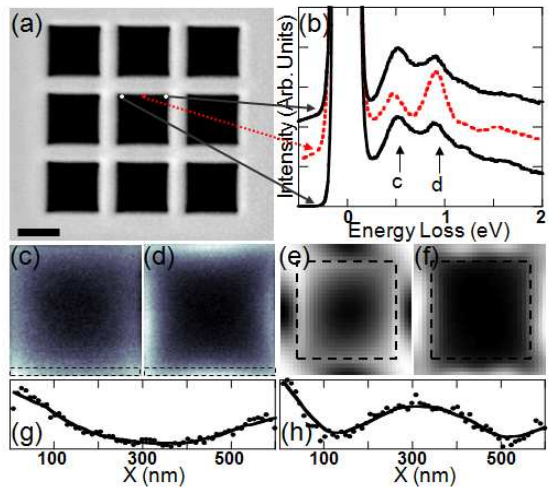

Fig. 3

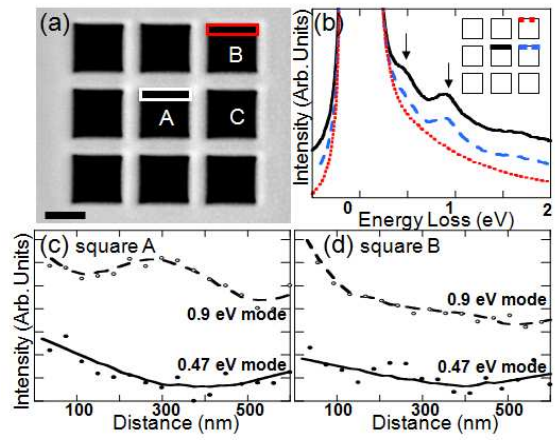

Fig. 4